\documentclass[namedreferences]{kluwer}

\usepackage{times}
\usepackage{mathptm}

\input{boxedeps}
\input{boxedeps.cfg}

\newcommand{\arcsec}{\hbox{$^{\prime\prime}$}}

\begin{document}

\begin{opening}
\title{Multi-Conjugate Adaptive Optics with two Deformable Mirrors - Requirements
and Performance}
\author{Thomas \surname{Berkefeld}\email{berke@kis.uni-freiburg.de}}
\institute{Kiepenheuer-Institut f\"{u}r Sonnenphysik, Freiburg, Germany}
\author{Andreas \surname{Glindemann}\email{aglindem@eso.org}}
\institute{European Southern Observatory, Garching, Germany}
\author{Stefan \surname{Hippler} \email{hippler@mpia-hd.mpg.de}}
\institute{Max-Planck-Institut f\"{u}r Astronomie, Heidelberg, Germany}

\begin{abstract}
In order to increase the corrected field of view of an adaptive optics (AO)
system, several deformable mirrors (DM) have to be placed in the conjugate
planes of the dominant turbulent layers (multi-conjugate adaptive optics, MCAO \cite{beckers}).
The performance of MCAO systems depends on the quality of the wavefront sensing of
the individual layers and on the number of corrected modes in each individual
layer as in single layer AO systems. In addition, the increase in corrected field
of view depends on the number of guide stars providing information about the
turbulence over a sufficiently large area in each turbulent layer.  In this paper,
we investigate these points and provide formulae for calculating the
increased field of view with a new approach using the spatial correlation
functions of the applied polynomials (e.g. Zernike). We also present a new scheme of 
measuring the individual wavefront distortion of each of the dominant
layers with a Shack-Hartmann-Curvature Sensor using gradient information as
well as scintillation. An example for the performance of a two layer MCAO system is
given for the 3.5-m telescope of the Calar Alto Observatory, Spain, using a
measured C$_n^2$-profile. The corrected field of view in K-band ($2.2\,\mu$m)
can be as large as 3 arcmin with a Strehl ratio above 60\%. 
\end{abstract}

\keywords{adaptive optics, multi-conjugate adaptive optics, laser guide stars, 
          turbulent layers}

\abbreviations{\abbrev{AO}{Adaptive Optics};
               \abbrev{MCAO}{Multi-Conjugate Adaptive Optics};
               \abbrev{DM}{Deformable Mirror};
               \abbrev{LGS}{Laser Guide Star};
               \abbrev{NGS}{Natural Guide Star};
	       \abbrev{GS}{Guide Star};
               \abbrev{SH}{Shack-Hartmann};
               \abbrev{SHC}{Shack-Hartmann-Curvature};
               \abbrev{NIR}{Near-Infrared}} 

\end{opening}

\section{Introduction}
\label{intro}

While AO systems increase the angular resolution of ground based telescopes  by a factor 
of 5--20, their most severe disadvantage is the very small corrected field of view which 
is typically of the order of 30\arcsec{} in K-band ($2.2\,\mu$m) and only a few arc seconds in
the visible. MCAO systems with several DMs in the conjugate planes of the dominant turbulent
layers as proposed by \inlinecite{beckers}  have two important advantages over conventional
AO systems:
\begin{itemize}
\item The corrected field of view is increased considerably by correcting the dominant 
turbulent layers instead of correcting the integrated wavefront aberrations even if the
turbulence is not located in distinct layers.
\item Focal anisoplanatism which is a severe limitation for Laser Guide Star (LGS) AO systems can 
be almost completely eliminated by using multiple LGSs (see Sect. 2.2).
\end{itemize}

The optimal position of the DMs is determined by the $C_n^2$-profile and by
the system characteristics e.g. the size of the
corrected field of view and the number of corrected modes. Usually the
angular anisoplanatism  of AO systems is calculated
according to the $(\theta / \theta_0)^{(5/3)}$-law \cite{tyson}. For real systems
however the corrected field of view is larger than given by
this formula. Therefore a more realistic calculation is needed, both for
estimating the performance of AO and MCAO systems,
i.e. the corrected FOV, and for obtaining the optimal position of the DMs
in MCAO systems.  This is done in section
\ref{erraniso}, where we derive an equation for calculating the
anisoplanatism both  for AO and MCAO systems using correlation
functions of the applied Zernike polynomials.  The anisoplanatism  depends
on the $C_n^2$-profile, the number of corrected modes
of each layer,  the number and position of the DMs and the angle between
science object and central guide star. Using various MCAO geometries, we apply these formulae to calculate the
Strehl ratios for a 3.5-m telescope and for a turbulence profile measured by Kl\"uckers \cite{klueckers}
at the Calar Alto Observatory. 

The second important issue of MCAO is the problem of  wavefront sensing
distinguishing the  turbulent layers. Section \ref{secSCI} gives a short introduction to
scintillation and describes the principle of separating the wavefront distortions of two turbulent
layers with a Shack-Hartmann (SH) sensor \cite{ribak2,schwartz}  or a
Shack-Hartmann-Curvature (SHC) sensor \cite{gli7}, both using intensity fluctuations. By
varying the effective altitude of the wavefront sensor(s) the  SNR of the scintillation signal can
be optimized. It will be shown in Section \ref{secSHC}  that a SHC sensor allows for much fainter
guide stars than a SH sensor. Furthermore, the SH sensor delivers only one scintillation signal,
therefore wavefront modes with zero-Laplacian cannot be properly measured. With a 
SHC sensor, however, the second scintillation signal can reduce the error induced by these modes.

A good introduction for the concept of MCAO can be found in \inlinecite{beckers}, more recent
investigations of its applicability and performance were presented by \inlinecite{johnston}. 

Unless mentioned otherwise, the wavelength used corresponds to the 
K-band ($2.2\,\mu$m).

\section{The wavefront error $\sigma^2_{\theta}$ of angular anisoplanatism}
\label{erraniso}
The wavefront error due to angular anisoplanatism is denoted by $\sigma^2_{\theta}$. It is 
caused by slightly different turbulence patterns in the light paths of two stars.
$\sigma^2_{\theta}$ can be expressed as

\begin{equation}
\sigma^2_{\theta} = \left( \frac{\theta}{\theta_0} \right) ^{5/3},
\end{equation}
where $\theta$ denotes the angular separation of the two stars. $\theta_0$ is the {\it isoplanatic
angle} that is defined such that $\sigma^2_{\theta} < 1$  for objects lying inside the 
isoplanatic patch \cite{tyson}. $\theta_0$ has to be defined in such a way that the number of
corrected modes is taken into account: A simple tip/tilt system has a  much larger isoplanatic
angle than higher order AO-systems.  For the special cases of a tip/tilt system and certain higher
order systems \inlinecite{sash} and \inlinecite{chassat} already derived expressions.

\subsection{Angular Anisoplanatism of AO-systems}
The following approach of calculating $\sigma^2_{\theta}$ is based on the cross-correlation 
function $C_{x,y}$ of two tip/tilt measurements at $b$ and at $b'$ and 
includes the general dependence of the isoplanatic patch on the degree of correction. 

Measurements of correlation functions yield important information about the parameters of 
AO systems: The temporal autocorrelation of tilt measurements determines the necessary 
bandwidth of the control loop and the correlation of tilt measurements of different stars (spatial
correlation) allows us to calculate of the isoplanatic angle.

The differential jitter of two beams caused by anisoplanatism is \cite{valley}

\begin{equation}
\langle(b-b')^2 \rangle_{x,y} = 2[1-C_{x,y}] \cdot \langle b^2 \rangle ,\label{ctosigmax} 
\end{equation}
where x- and y- directions correspond to the directions parallel and perpendicular to the line of
sight between the two stars. $\langle b^2\rangle$ represents the one dimensional variance of 
the tilt. 
The variance of the differential image motion can be written as

\begin{eqnarray}
(\Delta\theta)^2 &=& \langle (b-b')^2 \rangle_x+\langle (b-b')^2 \rangle_y  \nonumber \\ 
&=& \left((1 - C_x(\theta)) + (1 - C_y(\theta))\right) \cdot 2\langle b^2 \rangle,
\end{eqnarray}
where $2\langle b^2 \rangle$ is the two-axis variance of the position angle $b$. The 
variance $\langle b^2 \rangle$ of the position angle is related linearly to the wavefront error due
to G-tilt\footnote{The wavefront error related to the motion of the image centroid is often
called G-tilt since it is equivalent to the average gradient of the wavefront over the telescope
aperture.}. In a very similar fashion, one can write the wavefront error $\sigma^2_{\theta, {\rm
tt}}$ of the Zernike tilt - the first mode  (apart from piston) of the Zernike decomposition - due to
differential image jitter as

\begin{eqnarray} 
\sigma^2_{\theta, {\rm tt}} &=& \left((1 - C_{\rm tt, x}(\theta)) + (1 - C_{\rm tt,y}(\theta))\right) \cdot 2
\cdot 0.448\left(\frac{D}{r_0}\right)^{5/3} \nonumber \\ 
&=& 2\cdot \left(\frac{D}{r_0}\right)^{5/3} \cdot 0.448 \cdot (2-C_{\rm tt,x}(\theta)-C_{\rm tt,y}(\theta)).
\label{isotheta0} 
\end{eqnarray} 
$0.448 (D/r_0)^{5/3} $ is the wavefront variance of the single axis tilt as given by
\inlinecite{noll}. As soon as the correlation functions $C_{\rm tt,x,y}(\theta)$ have a value smaller
than 0.5 the  resulting tip-tilt correction with an on axis guide star deteriorates the image quality
since the wavefront error becomes larger than $0.448 (D/r_0)^{5/3}$, {\it i.e.} larger than the
uncorrected value. Eventually, for zero correlation the variance is twice as large as without
correction. 

The anisoplanatism of higher order wavefront modes can be deduced in a similar fashion
(see \inlinecite{diss}, corrected for minor inaccuracies in the deduction).
Then , the wavefront error $\sigma^2_{\theta,N}$ as a function of the
correlation of N corrected Zernike modes can be written as

\begin{equation} \sigma^2_{\theta,N} = 2\left(\frac{D}{r_0}\right)^{5/3} \cdot
\sum \limits_{j=1}^N (\sigma^2_{{\rm fit},j-1} - \sigma^2_{{\rm fit},j})(1-C_j(\theta)),
\label{isotheta} 
\end{equation}
summing up the different correlations $C_j(\theta)$ of each mode $j$ and the modal variance
$\sigma^2_{{\rm fit},j-1} - \sigma^2_{{\rm fit},j}$ to the wavefront error.  The total wavefront
error for a star at a distance $\theta$ from the guide star can then  be calculated by adding
$\sigma^2_{\theta,N}$ to the fitting error $\sigma^2_{fit,j>N}$.
Eq.~\ref{isotheta} can be used with
any set of polynomials if the correlation functions and the modal variances are available. 
\inlinecite{chassat} presented results for $\sigma^2_{\theta,N}$ for selected  Zernike-polynomials. 

Table ~\ref{wfehler} shows $\sigma^2_{\rm fit}$ for Kolmogorov turbulence 
described by Zernike polynomials as  a function of the number $j$ of corrected modes (see
\inlinecite{noll}). 

\renewcommand{\arraystretch}{1.5} \tabcolsep2mm 
\begin{table}[htb] 
\center{\begin{tabular}{|c|c|c|} \hline
$\sigma^2_{\rm fit,0}=1.0299 (D/r_0)^{5/3}$ & $\sigma^2_{\rm fit,7} \,\,=0.0525 (D/r_0)^{5/3}$ &
$\sigma^2_{\rm fit,14}=0.0279 (D/r_0)^{5/3}$ \\ $\sigma^2_{\rm fit,1}=0.5820 (D/r_0)^{5/3}$ &
$\sigma^2_{\rm fit,8} \,\,= 0.0463 (D/r_0)^{5/3}$ & $\sigma^2_{\rm fit,15}=0.0267 (D/r_0)^{5/3}$ \\
$\sigma^2_{\rm fit,2}=0.1340 (D/r_0)^{5/3}$ & $\sigma^2_{\rm fit,9}=0.0401 (D/r_0)^{5/3}$ &
$\sigma^2_{\rm fit,16}=0.0255 (D/r_0)^{5/3}$ \\ $\sigma^2_{\rm fit,3}=0.1110 (D/r_0)^{5/3}$ &
$\sigma^2_{\rm fit,10}=0.0377 (D/r_0)^{5/3}$ & $\sigma^2_{\rm fit,17}=0.0243 (D/r_0)^{5/3}$ \\
$\sigma^2_{\rm fit,4}=0.0880 (D/r_0)^{5/3}$ & $\sigma^2_{\rm fit,11}=0.0352 (D/r_0)^{5/3}$ &
$\sigma^2_{\rm fit,18}=0.0232 (D/r_0)^{5/3}$ \\ $\sigma^2_{\rm fit,5}=0.0648 (D/r_0)^{5/3}$ &
$\sigma^2_{\rm fit,12}=0.0328 (D/r_0)^{5/3}$ & $\sigma^2_{\rm fit,19}=0.0220 (D/r_0)^{5/3}$ \\
$\sigma^2_{\rm fit,6}=0.0587 (D/r_0)^{5/3}$ & $\sigma^2_{\rm fit,13}=0.0304 (D/r_0)^{5/3}$ &  \\ \hline
\end{tabular}} 
\caption{Wavefront error for Kolmogorov-turbulence described by Zernike polynomials as a
function of the number $j$ of corrected modes (Noll (1976)).}
\label{wfehler}
\end{table}

For $j>20$, $\sigma^2_{{\rm fit},j}$ can be approximated by

\begin{equation}
\sigma^2_{{\rm fit},j} \approx 0,2944\,(j+1)^{-\sqrt{3}/2}(D/r_0)^{5/3}.
\end{equation}
For an AO system with an LGS at infinity, i.e. without the cone effect, and an NGS for tip-tilt the
wavefront error $\sigma^2_{\theta, N, {\rm LGS}}$ becomes 

\begin{eqnarray}
\sigma^2_{\theta, N, {\rm LGS}} &=&
2\cdot\left(\frac{D}{r_0}\right)^{5/3}[0.448 \cdot
(2-C_x(\theta_{NGS})-C_y(\theta_{NGS}))+ \nonumber \\
&&\sum \limits_{j=3}^N (\sigma^2_{{\rm fit},j-1} - \sigma^2_{{\rm
fit},j})(1-C_j(\theta_{LGS}))].
\label{2laseriso}
\end{eqnarray}

\begin{figure}[htb]
\HideDisplacementBoxes
\centerline{\ForceWidth{100mm}
            \BoxedEPSF{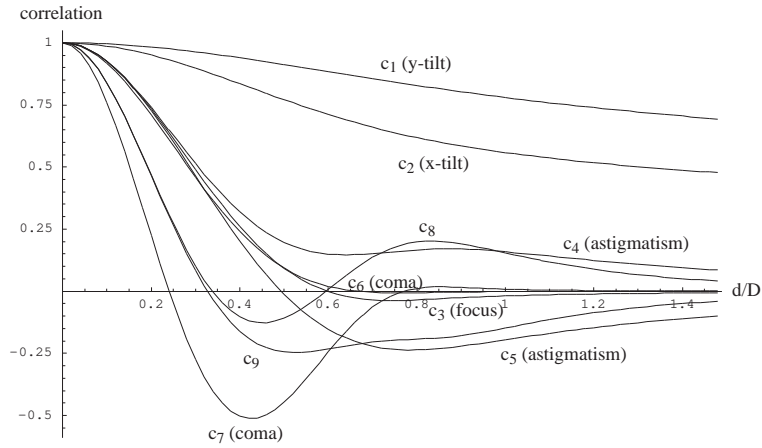}
           }\ForceOff
\caption{Correlation functions $c_j(d)$ for different Zernike-modes as a function of the light path
offset $d$ and the aperture $D$}
\label{tiltkorsp}
\end{figure}

The correlation functions $C_j(\theta)$ of the individual Zernike modes $j$ are
\begin{equation} 
C_j(\theta) = \frac{\int c_j(z\theta) C_n^2(z)\, dz }{\int C_n^2\,dz}, \label{2theokor} 
\end{equation} 
where $c_j(z\theta) = c_j(d)$ denotes the correlation caused by a single layer at altitude $z$ (\inlinecite{valley}). 
Figure \ref{tiltkorsp} shows the correlation for different aberrations as a function of the
parameter $d/D$.  Higher order aberrations generally decorrelate faster because they correspond to
smaller structures  (and thus smaller correlation lengths).

The turbulence profile $C_n^2$ determines the correlation functions $C_j(\theta)$ used in the
above equations. If $C_n^2$ is unknown, higher order correlations  can be approximately derived
from the measured tip/tilt correlations in pure Kolmogorov turbulence.

\begin{figure}[htb]
\HideDisplacementBoxes
\centerline{\ForceWidth{118mm}
            \BoxedEPSF{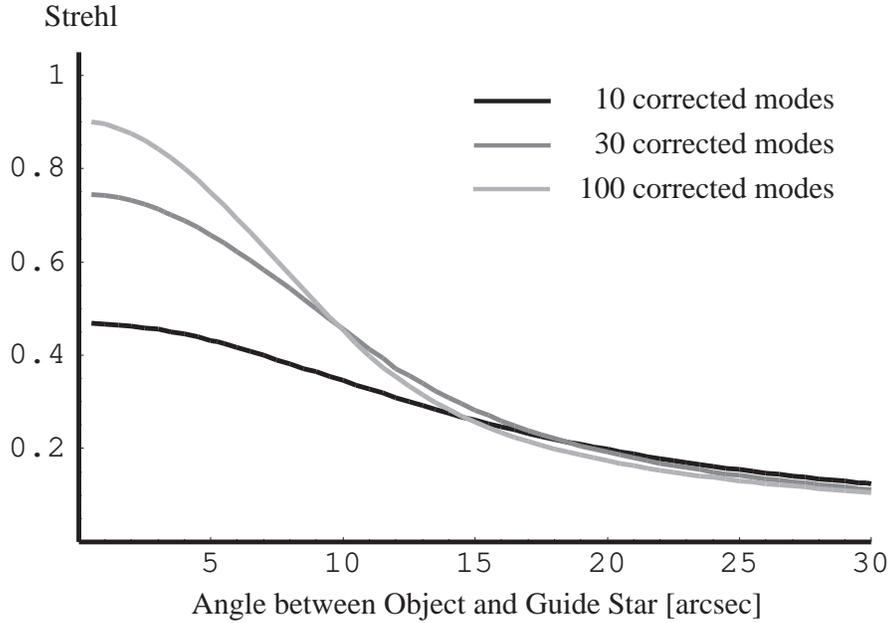}
           }\ForceOff
\caption{Strehl ratio for the wavefront variance $\sigma^2_{\rm fit}+\sigma^2_{\theta}$ as a
function of the number of corrected modes and the angle between science object and guide star ($D =
3.5$\,m, single turbulent layer at 5000\,m, $r_0 = 60$\,cm)}
\label{strehlwisomod}
\end{figure}

Inserting $C_j(\theta)$ into Eq.~\ref{isotheta} yields the wavefront variance
$\sigma^2_{\theta,N}$ for $N$ corrected modes. The isoplanatic angle $\theta_0$ can then be
calculated such that $\sigma^2_{\theta_0,N} = 1$. Fig.~\ref{strehlwisomod} shows the Strehl ratio
for the wavefront variance given by $\sigma^2_{\rm fit}+\sigma^2_{\theta,N}$  as a function of
the  number of corrected modes and the angle between science object and guide star.  Because of
the fast decorrelation of the higher modes, a low order correction can yield higher Strehl ratios
at large distances between object and guide star (see Fig. \ref{strehlwisomod}).

\subsection{Angular Anisoplanatism of MCAO-systems}
\begin{figure}[htb]
\HideDisplacementBoxes
\centerline{\ForceWidth{70mm}
            \BoxedEPSF{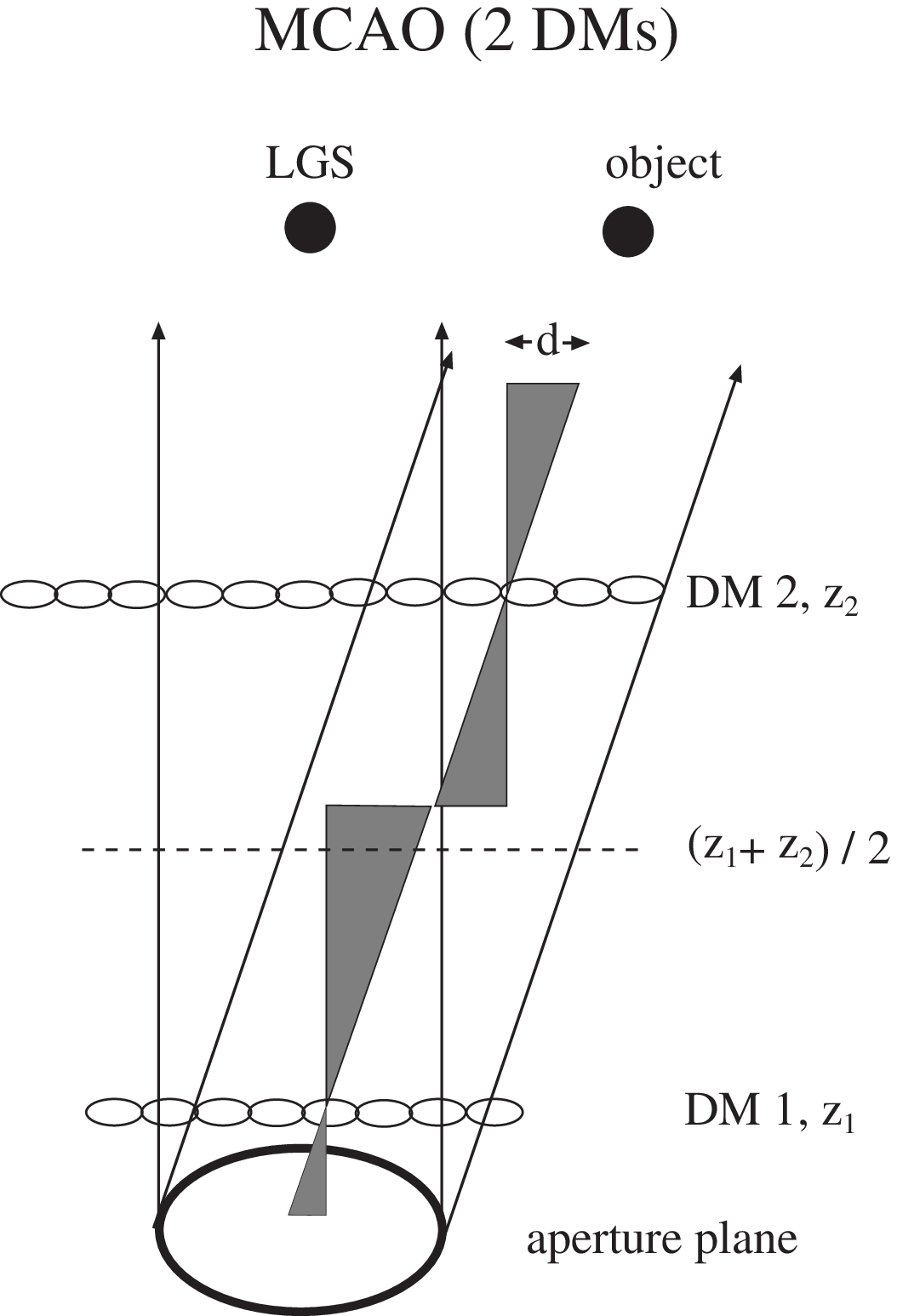}
           }\ForceOff
\caption{Anisoplanatism in an MCAO system with strong boundary layer turbulence: the shaded areas 
illustrate the amount of uncontrolled turbulence  between the DMs (darker shading = stronger
turbulence).  The width of the triangle equals the light path offset $d(z) = |z_{DM} -z|\theta$. The
anisoplanatism can be calculated using Eq.~(9), by integrating along the light path the correlation functions of
individual modes $c_j(d(z))$ weighted by the turbulence profile $C_n^2(z)$. }
\label{mcaoh}
\end{figure}

\begin{figure}[htb]
\HideDisplacementBoxes
\centerline{\ForceWidth{118mm}
            \BoxedEPSF{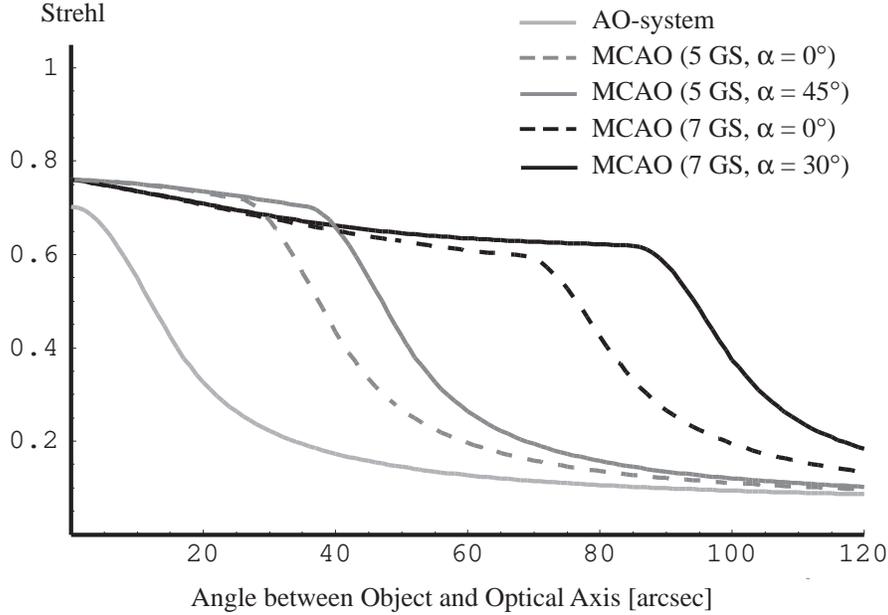}
           }\ForceOff
\caption{Strehl ratio for the wavefront variance $\sigma^2_{\rm fit}+\sigma^2_{\theta}$ 
as a function of the angle between object and central LGS
for a 2 DM  MCAO system and LGS geometries shown in Fig.~\ref{46eck} 
(see section \ref{sect:LGSG}). For comparison the
Strehl ratio is also given for a conventional AO system. The parameters are:  telescope aperture
$D = 3.5$\,m,  Calar Alto $C_n^2$-profile, K-Band and 37 corrected modes.  The Strehl ratio
depends both on the azimuth angle $\alpha$ and the distance to the central LGS.  The tip/tilt of
each LGS is assumed to be known.  The superiority of an MCAO system over an AO system, especially when
seven LGSs are used,  is obvious. The diameter of the corrected field of view can be as large as 3 arcmin. Due to
the corrected  {\it cone effect}, the MCAO system shows better performance even at the field center.}
\label{46eckstrehl}
\end{figure}

In the case of an MCAO system with $M$ DMs at conjugated altitudes $z_i$, the correlation 
$C_j(\theta)$ becomes
\begin{equation}
C_j(\theta) = \left( \sum\limits_{i = 1}^M \int\limits_{(z_{i-1}+z_i)/2}^{(z_i+z_{i+1})/2} c_j(|z_i-z|\theta) 
C_n^2(z) dz \right) \left/ \int\limits_0^{\infty} C_n^2(z) dz \right., 
\label{6cnneu}
\end{equation}
where $(z_0+z_1)/2 = 0$ (the integration starts from the ground) i.e. $z_0 := -z_1$, and $z_M+z_{M+1}$
equals the upper turbulence limit of the atmosphere.  For simplification it has been 
assumed that the number of corrected  modes per area is the same in all layers, i.e. DMs
correcting the high altitude turbulence have to correct more  modes (because of the larger
covered area) than the low altitude DMs.  Due to the small light path offset between object and
guide star in an MCAO system, the correlation functions have larger values than those of AO
systems (see Fig. ~\ref{mcaoh}). If there are more turbulent layers than DMs
Eq.~\ref{6cnneu} can be used to define the conjugate altitudes $z_i$ of the individual DMs by optimizing
$C_j(\theta)$. Inserting $C_j(\theta)$ into Eq.~\ref{isotheta} yields the angular
anisoplanatism $\sigma^2_{\theta}$ of an MCAO system (see Fig.~\ref{46eckstrehl}). If tip/tilt is measured by using
an NGS Eq.~\ref{2laseriso} has to be applied. 

\subsection{Laser Guide Star Geometry}
\label{sect:LGSG}

\begin{figure}[htb]
\HideDisplacementBoxes
\centerline{\ForceWidth{100mm}
            \BoxedEPSF{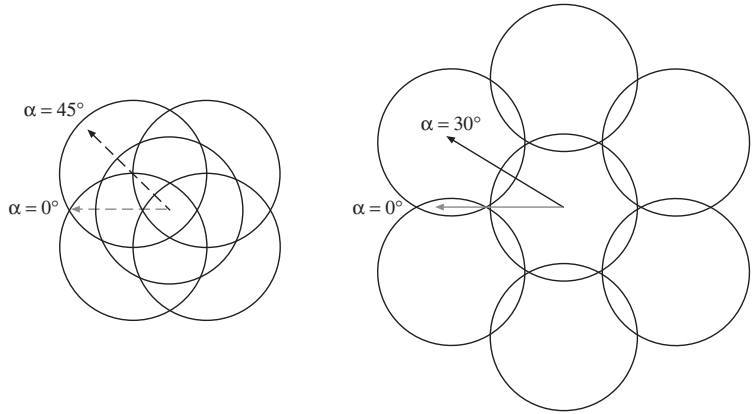}
           }\ForceOff
\caption{Top view at the highest turbulent layer. Five (left) and seven (right) LGSs have been
used for the following calculations. Each circle denotes the illuminated area of the
light cone from an LGS in the turbulent layer. If e.g. the Calar Alto turbulence profile is taken the Sodium layer
at 90\,km and the turbulent layer at 7\,km result in a illuminated circle with a diameter of 92\% of the
aperture of the telescope. These parameters are used throughout this paper.
For measuring the wavefront distortion over the desired field of view,  the light cones, i.e. the corresponding
subapertures of the wavefront sensors have  to cover the field of view completely. The field angles used for
Fig.~\ref{46eckstrehl} (0$^{\rm o}$, 30$^{\rm o}$, 45$^{\rm o}$) are also shown.} 
\label{46eck}
\end{figure}

To measure the wavefront distortion, several guide stars are necessary to cover the field.
Since it is unlikely to find natural guide stars at the desired  positions, LGSs will
be necessary in most cases. Mapping the wavefront distortion over the desired field requires the
LGSs to be pointed such  that their light cones still overlap at the altitude of the highest
turbulent layer (see Fig.~\ref{46eck}). 

Usually the science object does not coincide with an LGS, so different parts of the wavefront
intersect different LGS light cones and  are corrected accordingly. Therefore the anisoplanatism depends 
on the angular distance to different LGSs. In this paper, the fraction of intersection of each LGS is taken as a
weighting factor for its contribution to anisoplanatism. This approximation
leads to a linear Strehl ratio decrease  close to the field center, see Fig.~\ref{46eckstrehl} 
(instead of a more gaussian-like decrease, see Fig. ~\ref{strehlwisomod}). 
However, it is also possible to achieve an  almost
flat Strehl ratio over the FOV if the peak Strehl ratio at the field center is sacrificed for achieving
a higher Strehl ratio at the edge of the field. Fig.~\ref{46eckstrehl} shows the  Strehl ratio as a
function of the azimuth angle $\alpha$ and the angular distance between the object and
the central LGS. It is obvious that seven LGSs lead to a larger and more evenly covered field of 
view than five LGSs, although the Strehl  ratio drop-off due to residual angular anisoplanatism is
slightly higher. 

\begin{figure}[htb]
\HideDisplacementBoxes
\centerline{\ForceWidth{80mm}
            \BoxedEPSF{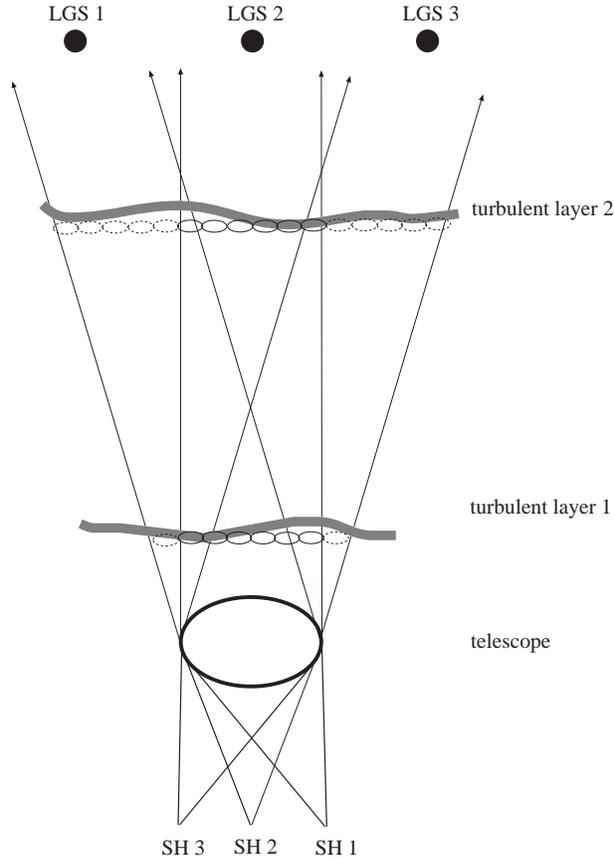}
           }\ForceOff
\caption{Side view of an MCAO geometry. Each circle in the turbulent layers indicates one
Shack-Hartmann subaperture i.e. one actuator of the DM that is situated in the conjugate
plane.  Solid and dashed circles denote the correction for the central LGS and the other LGSs,
respectively. At low altitudes, off-axis light cones mostly coincide with that of the (optimal
corrected) central LGS, whereas at high altitudes, an independent correction of off-axis
wavefronts is possible.}
\label{bereich}
\end{figure}

The positions of the LGSs correspond to the highest accuracy for wavefront sensing (as in
AO-systems, no angular anisoplanatism). At low altitudes, however, the  light cones of the LGSs
overlap almost completely. The wavefront reconstructions of the low layer are slightly different
for the individual LGSs because the extended turbulent layers are reduced to thin layers. Since
one has to choose one reconstruction to steer the mirror we used the central LGS, which defines
the center of the corrected FOV. The angular anisoplanatism of the other LGS  positions is mainly
due to this low altitude correction.  For high altitudes and little overlap between neighbouring
light cones, more degrees of  freedom allow a correction close to the optimum, as can be seen in
Fig.~\ref{bereich}. A more even correction  can be achieved when the low altitude DM corrects
the low altitude {\it average} wavefront error of all  LGS directions. This, however, results in a
slight loss of  Strehl ratio on-axis. For this paper, we chose the optimal correction of
the field center.

\begin{figure}[htb]
\HideDisplacementBoxes
\centerline{\ForceWidth{118mm}
            \BoxedEPSF{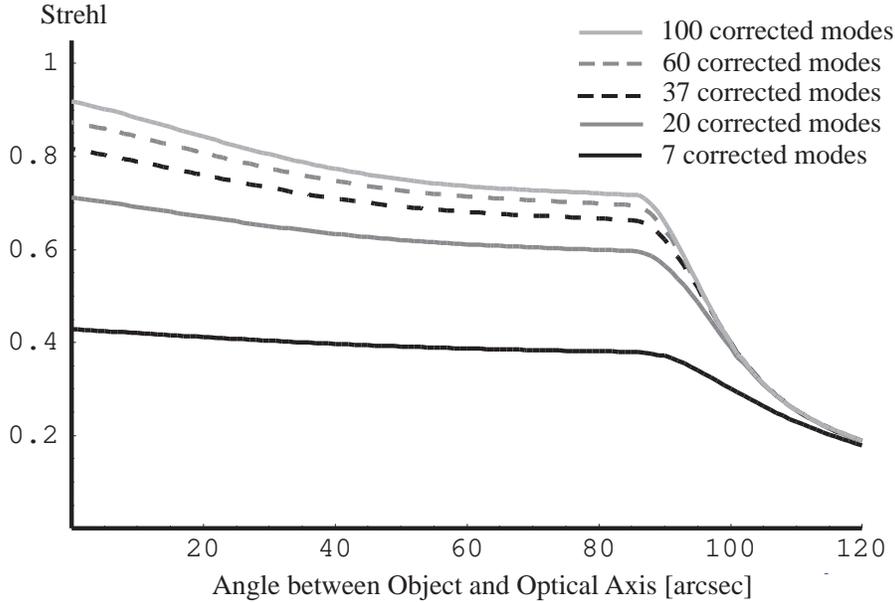}
           }\ForceOff
\caption{Strehl ratio for the wavefront variance $\sigma^2_{\rm fit}+\sigma^2_{\theta}$ of an MCAO (2 DMs) as
a  function of the number of corrected modes and the angle between  science object and optical axis ($D =
3.5$\,m, 37 corrected modes, Calar  Alto $C_n^2$-profile, K-Band, 7 LGS, $\alpha = 30^{\circ}$  
(see Fig.~\ref{46eck})).}
\label{mcaowisomod}
\end{figure}

\begin{figure}[htb]
\HideDisplacementBoxes
\centerline{\ForceWidth{118mm}
            \BoxedEPSF{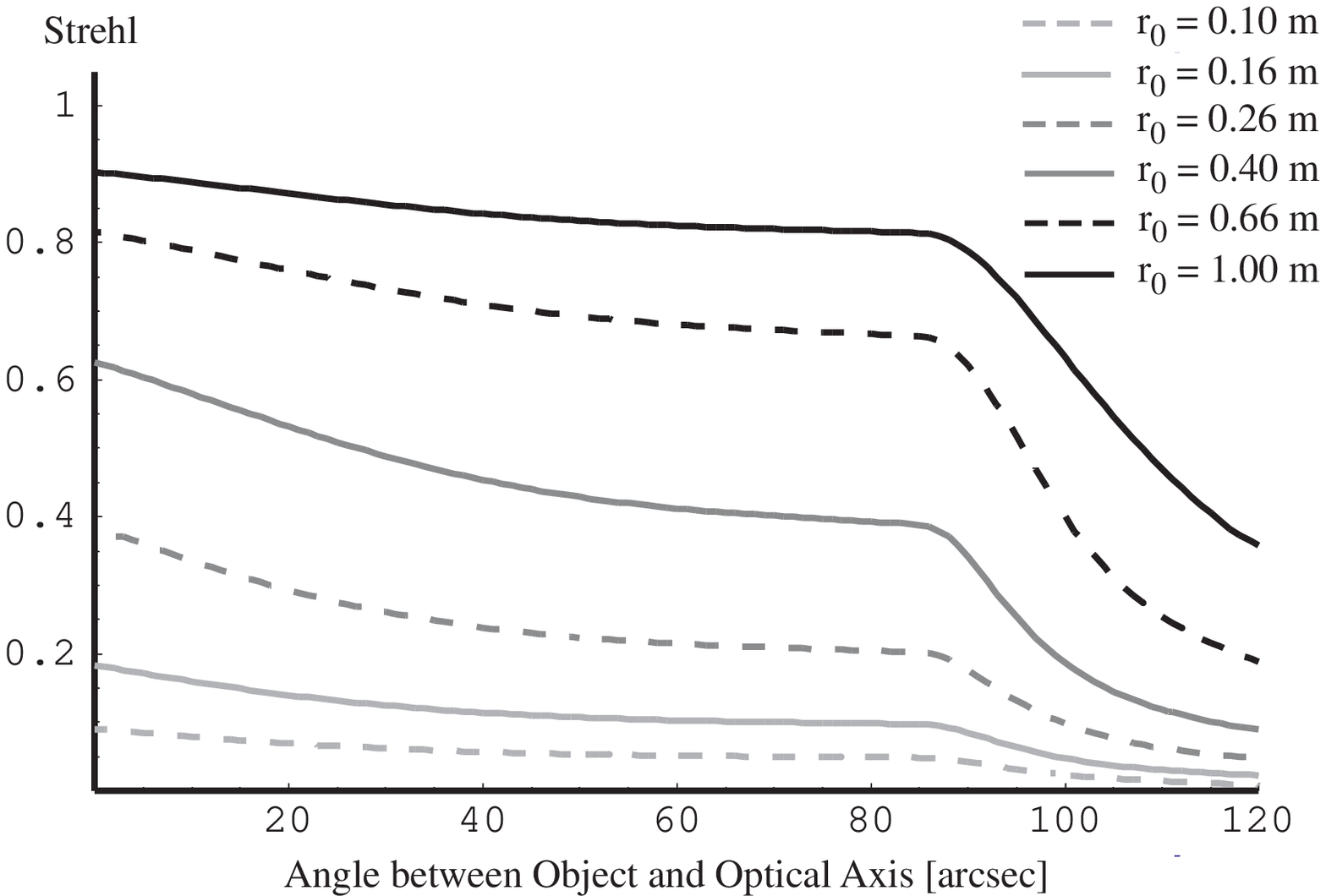}
           }\ForceOff
\caption{Strehl ratio for the wavefront variance $\sigma^2_{\rm fit}+\sigma^2_{\theta}$ of an MCAO (2 DMs) as
a  function of $r_0$ (scaled Calar Alto $C_n^2$-profile) and the angle between  science object and optical axis
($D = 3.5$\,m, 37 corrected modes, 7 LGS, $\alpha = 30^{\circ}$ (see Fig.~\ref{46eck})).}
\label{mcaowisor0}
\end{figure}

The Strehl ratio for the wavefront variance $\sigma^2_{\rm fit}+\sigma^2_{\theta}$ of an MCAO system with 2
DMs is shown as a function of the angle between science object and guide star for different numbers of
corrected modes (Fig.~\ref{mcaowisomod}) and different $r_0$ (Fig.~\ref{mcaowisor0}) . It
is obvious that a high order correction at large angles $\theta$ results in a higher gain for the
Strehl ratio in MCAO than in single DM AO systems.  For our calculations we assumed that the
absolute tip/tilt of the LGSs is known (see end of section 2).  Due to the $(D/r_0)^{(5/3)}$-law (see
Eq.~\ref{isotheta}), both AO and MCAO systems heavily depend on  good seeing to reach a high
image quality. 

Other possibilities of calculating the isoplanatic angle of MCAO systems  are given by
\inlinecite{toko} and \inlinecite{wallner2}. Their solutions, however, do not take the number of
corrected modes into account. 

A number of methods have been proposed to deal with the tip/tilt determination
problem. Although it is possible to use only one natural guide star (NGS) for absolute tip/tilt
determination  over the whole field of view \cite{johnston}, the wavefront reconstruction would not
be as accurate as if the absolute tip/tilt was determined for each LGS.  Since it is unlikely to find
one NGS close to each LGS, the absolute tip/tilt of each LGS should be measured independently. 
\inlinecite{raga} proposed to position and point auxiliary telescopes in such a way that
one LGS and one NGS are in a line with one auxiliary telescope. Then the tip/tilt of each LGS can
be determined by subtraction\footnote{Since one auxiliary telescope can measure  tip {\it or} tilt,
two auxiliary telescopes per LGS are required for absolute  tip/tilt-determination.}. 
Another possibility is the use of polychromatic LGSs \cite{poly1}, which excite two
wavelengths in the sodium layer. The wavelength difference of the two colours leads to  a
differential tip/tilt caused by atmospheric dispersion, from which the absolute tip/tilt can be
calculated.  Unfortunately, for a sufficient excitation of the second colour, the laser
output power has to be increased by about two orders  of magnitude.

\section{Separating the wavefront distortion of high and low altitude turbulent layers}
\label{separr}

It has been shown in the previous section that 2-DM-MCAO systems already produce a wide
and relatively evenly corrected  field of view. In this section, we discuss methods to distinguish
the wavefront aberrations of the corresponding two layers, a high and a low altitude layer. 

One possibility is the use of tomographic methods, as proposed by \inlinecite{beckers}. Since a
large number of  LGSs is required, this method is quite demanding for night telescopes, but
can be very useful for solar telescopes where no LGSs are needed. New aspects of zonal
tomography can be found in \inlinecite{tf90}. The possibility of  modal tomography has
recently been shown by Ragazzoni \cite{raga2}\cite{raga4}. Solar AO systems  were discussed by
\inlinecite{scharmer} and \inlinecite{rimmele}. 

In order to reduce the number of LGSs\footnote{However, the desired FOV should be 
completely covered by the LGS light cones.} for measuring the individual wavefront distortion of
two turbulent layers, the  intensity information provided in each lenslet of a SH
sensor can be used, as proposed by \inlinecite{ribak}.If the SH sensor is situated in 
the conjugate pupil plane, the intensity variation at low light
levels is dominated by photon noise. This situation can be improved by varying the position
of the SH sensor as will be discussed in the following. For further SNR improvement and for 
reducing the error caused by the wavefront modes undetectable by scintillation, a second
SH sensor can be used, effectively forming the  SH-Curvature sensor. 

Further aspects of wavefront separation and reconstruction can be found in 
\cite{ro2,hickson,ribak2,schwartz}.

\subsection{Introduction to scintillation}
\label{secSCI}
\begin{figure}[htb]
\HideDisplacementBoxes
\centerline{\ForceWidth{90mm}
            \BoxedEPSF{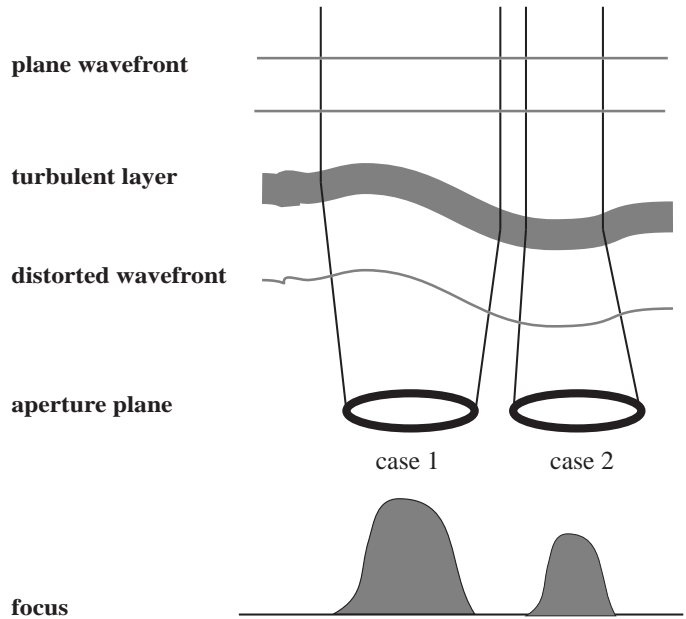}
           }\ForceOff
\caption{A turbulent layer focuses (case 1) or defocuses (case 2) the incoming light, resulting in
a larger or smaller effective aperture in the focal plane.} 
\label{szi}
\end{figure}

Intensity fluctuations of star images (scintillation) are caused by the curvature of turbulent 
layers (second derivative of the phase, lensing effect, as shown in Fig.~\ref{szi}).

Usually, astronomical observations are not affected by scintillation due to long
integration times and large apertures. However, if intensity fluctuations of highly time-resolved 
measurements are  not treated as noise, they can be used to provide information about the
altitude distribution of  turbulent layers. Good introductions to the theory of scintillation can be
found in \cite{dravins1,dravins2,dravins3,elsaesser,reiger}. 

The variance of the measured intensity $\sigma^2_I$ consists of the variance of the  scintillation
$\sigma^2_S$ and other noise sources (detector noise $\sigma^2_D$ and  photon noise
$\sigma^2_P$):

\begin{equation}
\sigma^2_I = \sigma^2_S + \sigma^2_D + \sigma^2_P.
\end{equation}
If the detector noise and the photon noise are known, the variance of the scintillation can be
determined. According to \inlinecite{reiger}, the theoretical value of $\sigma^2_S$ is given by

\begin{equation} 
\sigma^2_S \propto D^{-7/3}(\cos\gamma)^{-3} \int z^2 C_n^2(z)\, dz, 
\label{szivar} 
\end{equation} 
$D$ and $\gamma$ denoting the telescope aperture and the zenith angle, respectively.
Simulations with  the program {\it Turbulence} \cite{berke2} have shown that the normalized
variance of the intensity for a single layer at altitude $z$ can be approximated by

\begin{equation} 
\sigma^2_S = D^{-7/3}\lambda^2z^2r_0^{-5/3}(\cos\gamma)^{-3}. 
\label{szi2r0} 
\end{equation}
Because of $r_0 \propto \lambda^{6/5}$,  the scintillation does not depend on the wavelength. 
For an extended turbulence profile one obtains

\begin{equation} 
\sigma^2_S = 16.7\cdot D^{-7/3}(\cos\gamma)^{-3}\int z^2 C_n^2(z)dz. 
\label{szi2cn} 
\end{equation}
Due to the factor $z^2$ the scintillation is mostly caused by high altitude turbulent layers. 
For $\sigma_S > 10\%$, e.g. for large zenith angles $\gamma$ or for small telescope apertures $D$, 
the  scintillation begins to become nonlinear, approaching a maximum value, and Eq.~\ref{szi2cn}
is no longer valid \cite{dravins1,protheroe}.

\subsection{Wavefront sensors for separating the contributions of high and low altitude
turbulent layers}
\label{trennung}

\begin{figure}[htb]
\HideDisplacementBoxes
\centerline{\ForceWidth{90mm}
            \BoxedEPSF{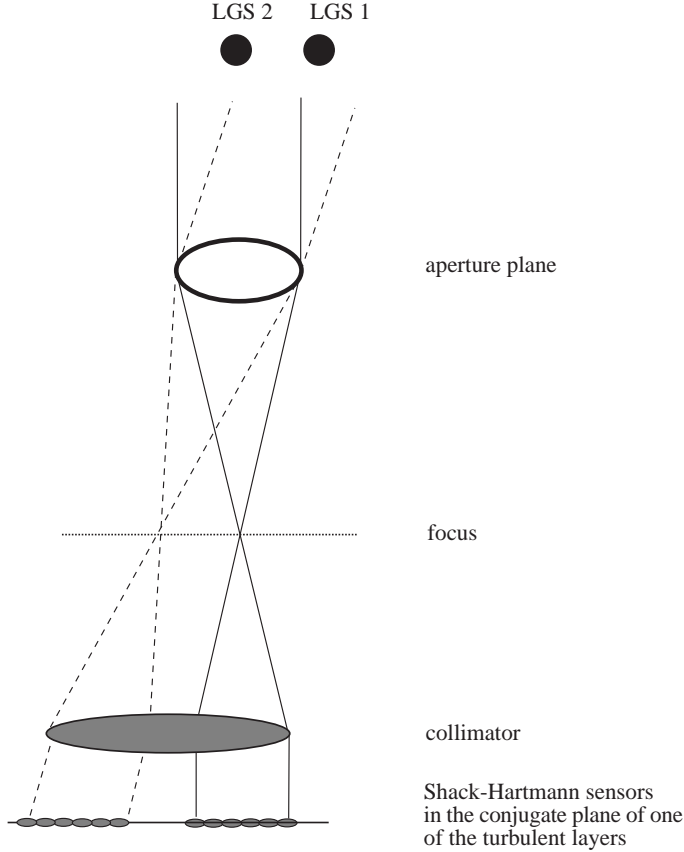}
           }\ForceOff
\caption{Setup of a wavefront sensor separating the wavefront distortions  of two turbulent
layers. For clarity, the light paths of only two guide stars have been plotted. One SH sensor per
LGS is  required.}
\label{6trennung}
\end{figure}

Since the wavefront gradient in each SH subaperture is the {\it sum} 
of the wavefront distortions of all turbulent layers, additional information is needed for
distinguishing the influence of individual layers.  By using the spatially resolved scintillation
provided in each subaperture of the SH sensor, the distortion of one of the layers can be
reconstructed. Together with the  known sum of the aberrations, the wavefront 
distortion of the other layer can then also be determined (Fig.~\ref{6trennung}).

As in a conventional AO-setup, the SH sensor measures the tip/tilt in each subaperture,  thus
delivering the integrated wavefront error. Because the SH sensor is situated in the  conjugate
plane of one of the turbulent layers (subsequently named {\rm first} layer), this layer has no 
effect on the intensity of the SH pattern.   The - for the SH sensor - defocused other
(subsequently named {\rm second}) layer leads to intensity fluctuations $I_2' (x_2')$ in the 
subapertures from which the wavefront distortion $\phi$ of the second layer can be
reconstructed \cite{ribak,gli7}: 

\begin{equation}
I_2' (x_2') = 1 - \frac{z_{2}'-z_{1}'}{2k} \frac{\partial^2 \phi_{1} (x_{2}')}{\partial x'^2_2},
\end{equation}
where indices denote the layers. $z$ is the altitude coordinate, $k = 2\pi/\lambda$, and
dashed variables indicate the  conjugate planes. The radial wavefront tilt affects the intensity
measurement at the edge of the aperture \cite{ro2}.  Although the edge radial tilts can be  
extracted directly from the SH tilt measurement and thus their contribution can easily be
corrected, they must  be known for each layer.  They can be  estimated by first determining the
mean gradient $m_i$ (averaged over all LGSs $j$) of each edge subaperture $i$. The $m_i$ can be 
regarded as the edge gradients of the low altitude turbulence (which are the same for all LGSs),
whereas its deviation $r_{ij}$ can be  regarded as the high altitude gradients

\begin{equation}
r_{ij} = t_{ij}-m_i  \qquad {\rm with} \quad m_i = \frac{\sum\limits_{j=1}^n t_{ij}}{n},
\label{radtilt}
\end{equation}
with $t_{ij}$ being the measured gradients. It is obvious that this is only an approximation
which will be more accurate with a higher number of LGS (and thus a larger corrected field
of view). Once the wavefront distortion of one of the layers has been  reconstructed,
subtraction from the measured sum of the wavefront errors delivers the wavefront distortion
of the other layer.  The error made by averaging the radial gradients, as described by
Eq. \ref{radtilt}, is still unknown. This problem should therefore be  addressed more closely in the future.

One can define the contribution of an individual layer $i$ to the scintillation in a conjugate plane
of altitude $h$ by (see Eq.~\ref{szi2cn})
\begin{equation}
\sigma_{S, i}^2(h) = 16.7\cdot
D_{\rm sub}^{-7/3}(\cos\gamma)^{-3}\int\limits_{\rm L_i} (z-h)^2 C_n^2(z)dz,
\end{equation}
with $D_{\rm sub}$ the subaperture size of the SH sensor that is placed in the conjugate plane of
altitude $h$.

The goal of the measurement is to determine the contribution of the more turbulent of the two
layers. Thus, the  SH sensor is placed in a conjugate plane where the scintillation of the
more turbulent layer (in this case layer 2) is very large and the contribution of layer 1 is
very small. Then, one can write the signal-to-noise ratio (SNR) as the normalized mean square
error (1/SNR$^2$) of the intensity measurement as:

\begin{equation}
\frac{1}{{\rm SNR}^2} = \frac{\frac{1}{N_{\gamma}}+\frac{n\cdot R^2}{N_{\gamma}^2}+ 
\sigma_{S, 1}^2(h)}{\sigma_{S, 2}^2(h)} . 
\label{6szifehler}
\end{equation}
The numerator consists of the photon noise $1/N^{1/2}$, the read noise $R\cdot n^{1/2}$ ($n$
is the number of pixels used  for the intensity measurements, $R$ denotes the readout noise
per pixel), and the contribution of layer 1 that is treated as noise for the measurement of
layer 2 that constitutes the signal in the denominator.  The scintillation error due to the non
discrete layering of the $C_n^2$-profile is very small and need not be taken into account \cite{diss}. 

\begin{figure}[htb]
\HideDisplacementBoxes
\centerline{\ForceWidth{100mm}
            \BoxedEPSF{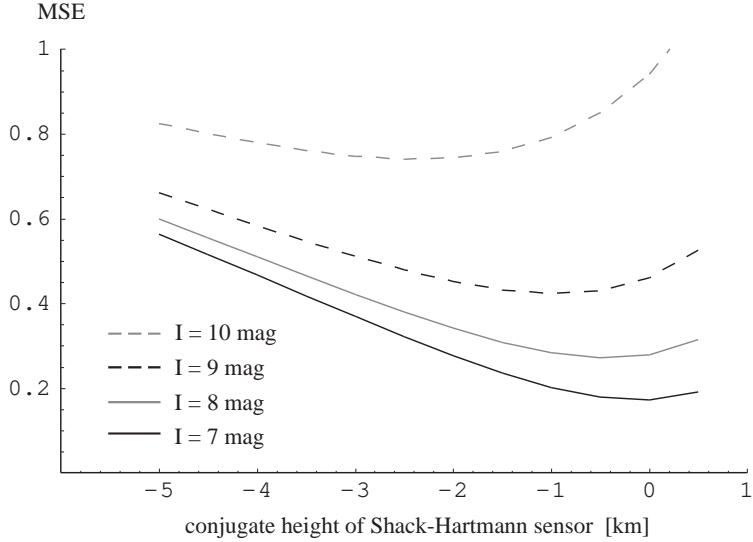}
           }\ForceOff
\caption{Normalized mean square error (1/SNR$^2$) of the scintillation measurements as a
function of the  conjugate altitude of the SH sensor for  different LGS intensities. The subaperture 
size and the read noise have been assumed as $D_{\rm sub} = 0.5$\,m (35-40 corrected modes) 
$n = 16$ pixels and R = 3e$^-$, which are typical values for the AO system ALFA at the 
Calar Alto 3.5\,m telescope.  The integration time of
the wavefront sensor was chosen to be optimal for each  LGS brightness. In this figure, the
scintillation caused by the high altitude turbulence is measured.  Increasing LGS brightness leads
to increasing optimal altitude of the SH sensor. In order to have an SNR better than 4 the mean
square error has to be smaller than 1/16.} 
\label{sziopt}
\end{figure}

Typically the measured scintillation is of the order of a few percent. This poses problems at low 
light levels where the scintillation competes with the shot noise. Although it is possible to
measure the scintillation of the more  turbulent layer (usually the ground layer) and thus get a
smaller relative error, this does not lead to a better separation of the  layers because the {\it
absolute} error of the wavefront reconstruction is also proportional to the turbulence strength. 
Therefore it does not matter whether the scintillation of the strong or the weak turbulence is
measured. Instead, the mean square error can be reduced by moving the SH sensor further
away from the conjugate plane of the first turbulent layer,  as shown in Fig.~\ref{sziopt}.

Although this leads to an unwanted contribution to scintillation from the now defocused first 
layer, the SNR increases within certain limits due to the much higher scintillation signal from
the second layer.  However, in the case of a 589\,nm LGS with a 10 mag G-star brightness
equivalent, the maximum SNR  is 1.15, which is not sufficient for wavefront separation.

\subsection{The Shack-Hartmann-Curvature sensor}
\label{secSHC}
For low light levels the amount of scintillation has to be increased by moving the SH sensor 
further away from the conjugate planes of the turbulent layers. Since the scintillation effects of
two layers cannot be distinguished with  one SH sensor, an additional SH sensor is necessary,
forming the Shack-Hartmann-Curvature sensor.  In order to refine the SHC setup
presented by \inlinecite{gli7} we propose to position the individual SH sensors in such a way
that a mean intensity fluctuation of 10\% (the non-linearity limit of scintillation) is achieved in
each SH sensor. This is the case  at the conjugate planes of a very large positive altitude and a
large  negative altitude, leading to an excess of illumination in one plane and to a lack of
illumination in the other,  similar to the Curvature Sensor \cite{ro2}. For the Calar Alto
$C_n^2$-profile this results in altitudes of  approx. 16\,km and -12\,km. 

\begin{figure}[htb]
\HideDisplacementBoxes
\centerline{\ForceWidth{100mm}
            \BoxedEPSF{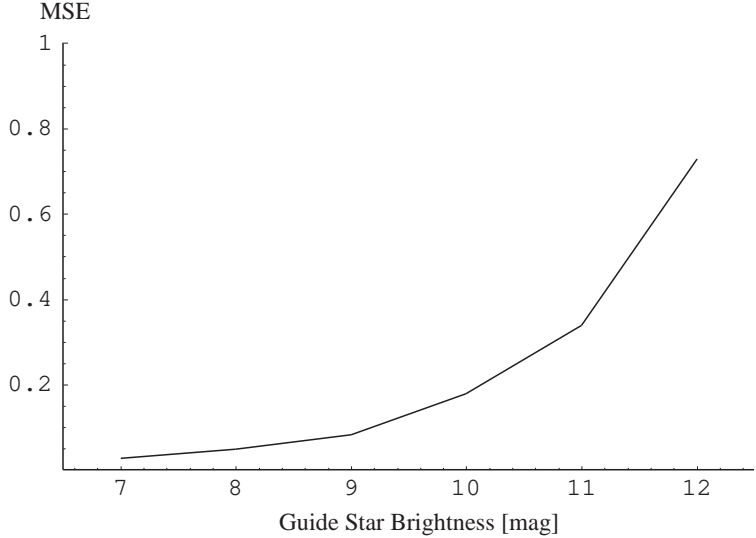}
           }\ForceOff
\caption{Normalized mean square error of the intensity measurements performed by a 
SHC sensor as a function of the  guide star brightness. The conjugate altitudes of the SH sensors
have been assumed as 16\,km and -12\,km, the other parameters are  the same as in fig
\ref{sziopt}.} 
\label{szifehler2}
\end{figure} 

The scintillation variance $\sigma_{S, {\rm SHC}}^2(h_1, h_2)$ of a SHC sensor can be expressed as
the sum of two SH sensor signals 
each of which receives contributions of high and low altitude turbulence. For turbulence separation, the 
signal of each SH sensor is the difference of high and low altitude turbulence scintillation:
\begin{eqnarray}
\sigma_{S, {\rm SHC}}^2(h_1, h_2) & = & 16.7\cdot D_{\rm sub}^{-7/3} (\cos\gamma)^{-3} \times \\
( \int\limits_{\rm L_1} ((z - h_1)^2 + (z - h_2)^2) C_n^2(z) dz & - & \int\limits_{\rm L_2}
((z - h_1)^2 + (z - h_2)^2) C_n^2(z) dz ) \nonumber
\end{eqnarray}

Then, the mean square error for a SHC is:
\begin{equation}
1/{{\rm SNR}^2} =  \frac{ \frac{1}{N_{\gamma}}+ \frac{n\cdot R^2}{N_{\gamma}^2}}
  {\sigma_{S, {\rm SHC}}^2(h_1, h_2)}
\label{6szifehler2}
\end{equation}
An LGS brightness of 8 mag leads to an SNR of 4 which should be sufficient for most
applications (see Fig.~\ref{szifehler2}).  Therefore observations  in the NIR and a typical LGS
brightness of 10\,mag require one SHC sensor per LGS. Observations in the visible require
much more powerful LGSs in order to sense and correct a higher number of modes.  Then, a
single SH per LGS can be used. In any case, the $C_n^2$-profile and the zenith angle  should be
monitored continuously in order to adjust the SH sensors and the DMs accordingly.

There is another aspect that makes the use of a SHC sensor more desirable than a SH sensor: 
Since the scintillation signals of the wavefront modes are not linearly independent, it is not
possible to measure the wavefront modes of zero Laplacian with a SH sensor directly. Although it is 
possible to estimate these modes by averaging the radial wavefront tilts as described in section \ref{trennung}, 
the remaining error should be reduced further by using the intensity measurements 
for applying a priori (statistical) knowledge. Since a SHC sensor provides two and a SH sensor only one 
measurements per subaperture, the application of a priori knowledge will work better with a SHC sensor. 
However, for deriving a stable algorithm that operates in real-time, 
more detailed work has to be done in the field of wavefront reconstruction by intensity measurements.  

\section{Conclusion}
\label{conclusion}

By using MCAO, it is possible to overcome the most severe disadvantage of AO,  the very small
corrected FOV.   We have shown how the geometry of MCAO systems affects the angular
anisoplanatism and thus the  size of the corrected FOV. A setup with seven LGSs for wavefront
sensing leads to a wide and relatively evenly corrected field.  In the case of the Calar Alto 3.5\,m 
telescope, this setup would lead to a FOV of about three arcminutes. Eq.~\ref{isotheta} plays an
important role in calculating the wavefront error caused by angular anisoplanatism. By its
minimization one obtains the optimal position of the DMs (according to  the $C_n^2$-profile).
Since the remaining anisoplanatism inside the field is rather small, a wavefront correction with
two DMs seems to be a good compromise between anisoplanatism, cost  and system
complexity. The separation of the wavefront errors of the two layers can be 
accomplished in various ways: For solar telescopes, tomographic methods, as proposed by
Beckers, seem to be the most accurate and  easy to implement way. For night telescopes, the
separation can be done by using the intensity  information provided by each lenslet of the SH
sensor or SHC sensor. The SHC sensor should be preferred because it allows a  fainter limiting
magnitude and can reduce the separation error caused by the wavefront modes with zero
Laplace operator.

The high costs for the LGS setup and the absolute tip/tilt determination will prevent MCAO
systems from being used at existing 3.5\,m-class-telescopes, at least as long as other 
improvements of conventional AO systems are possible. At 8+\,m-class-telescopes however,
focal anisoplanatism will  require the use of multiple laser guide stars. Furthermore, the cost of
MCAO systems compared  to those of the telescope will decrease, so that MCAO will become a
common feature at large telescopes.

\begin{appendix}

\section{Calar Alto $C_n^2$-profile}
\label{cncaha}

\begin{figure}[htb]
\HideDisplacementBoxes
\centerline{\ForceWidth{80mm}
            \BoxedEPSF{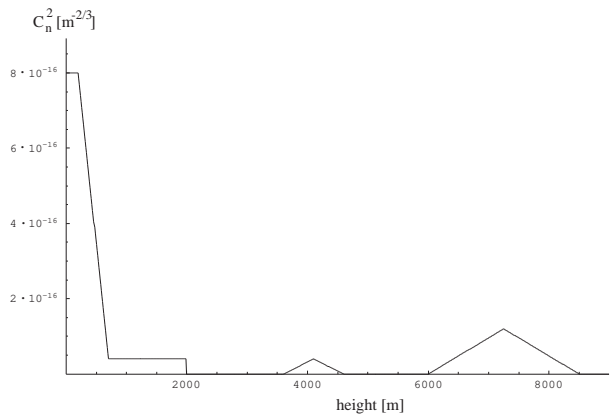}
           }\ForceOff
\caption{$C_n^2$-profile at the Calar Alto Observatory 
         (Klueckers et al., 1998),
         $r_0 = 66$\,cm in K-band)} 
\label{cnneu}
\end{figure}

Fig.~\ref{cnneu} shows a slightly simplified $C_n^2$-profile measured by 
\inlinecite{klueckers} at the Calar Alto Observatory, Spain. The upper turbulent layer at an
altitude of 7\,km delivers the main contribution to the angular  anisoplanatism, the lower turbulent layer
determines most of $r_0$. For a 2-DM-MCAO, the optimal altitudes for the DMs are 400\,m and 6900\,m.
It should be noted, that the turbulence profile can change  rather quickly. Kl\"uckers reported a
change of the upper turbulent layer strength by a factor  of two in only a few minutes.
Therefore, frequent $C_n^2$-measurements should be made to adjust the optimum altitude of the
DMs.

\end{appendix}


\bibliographystyle{klunamed} 
\bibliography{MCAOrefs} 

\end{document}